\documentstyle[preprint,aps,eqsecnum]{revtex}
\def\beq#1{\begin{equation} \label{#1}}
\def\eeq{\end{equation}}
\def\bra#1{\left\langle #1\right\vert}
\def\ket#1{\left\vert #1\right\rangle}

\def\PLB{{ Phys. Lett.} B}

\def\PRD{{ Phys. Rev.} D}
\begin{document}
{
\tighten
\title{Quantum Mechanics of Neutrino Oscillations - Hand Waving for
Pedestrians}
\author{Harry J. Lipkin\,\thanks{Supported
in part by The German-Israeli Foundation for Scientific Research and
Development (GIF) and by the U.S. Department
of Energy, Division of High Energy Physics, Contract W-31-109-ENG-38.}}
\address{ \vbox{\vskip 0.truecm}
  Department of Particle Physics
  Weizmann Institute of Science, Rehovot 76100, Israel \\
\vbox{\vskip 0.truecm}
School of Physics and Astronomy,
Raymond and Beverly Sackler Faculty of Exact Sciences,
Tel Aviv University, Tel Aviv, Israel  \\
\vbox{\vskip 0.truecm}
High Energy Physics Division, Argonne National Laboratory,
Argonne, IL 60439-4815, USA\\
~\\HJL@axp1.hep.anl.gov
\\~\\
}
 
\maketitle
 
\begin{abstract}
 
  Why Hand Waving? All calculations in books describe oscillations in
time. But real experiments don't measure time. Hand waving is used to convert
the results of a ``gedanken time experiment" to the result of a real experiment
measuring oscillations in space. Right hand waving gives the right answer;
wrong hand waving gives the wrong answer. Many papers use wrong handwaving to
get wrong answers. This talk explains how to do it right and also answers the
following questions:
 
  1. A neutrino which is a mixture of two mass eigenstates is emitted with
muon in the decay of a pion at rest. This is a a ``missing mass experiment"
where the muon energy determines the neutrino mass. Why are the two mass states
coherent?
 
  2. A neutrino which is a mixture of two mass eigenstates is emitted at time
t=0. The two mass eigenstates move with different velocities and arrive at the
detector at different times. Why are the two mass states coherent?
 
  3. A neutrino is a mixture of two overlapping wave packets with
different masses moving with different velocities. Will the wave packets
eventually separate? If yes, when?
 
\end{abstract}
 
} 
 
\section {Introduction}
 
\subsection{History and Dedication}
 
This analysis of the basic physics of flavor oscillations began in 1981, when
Israel Dostrovsky, then working on the gallium-germanium chemistry for a  solar
neutrino experiment, invited me to give a series of talks at Brookhaven in a
language that chemists would understand. The notes of these
lectures\cite{Dost} were later expanded into lecture notes notes for a course
in quantum mechanics\cite{QM} and then given further in a talk at a GALLEX
collaboration meeting\cite{NeutHJL}. Meanwhile the gallium collaboration moved
to Grand Sasso to becme GALLEX. Dostrovsky has continued as one of the
leaders in the collaboration, while his pioneering chemistry developed for 
the separation and detection of tiny numbers of germanium atoms produced by
neutrinos in tons of gallium has been used by both GALLEX and SAGE.
 
It is a pleasure to dedicate this talk to my friend and colleague Israel
Dostrovsky on the occasion of his 80th birthday.
 
\subsection{Problems in the description and treatment of flavor oscillations}
 
Flavor oscillations are observed when a source creates a particle
which is a mixture of two or more mass eigenstates, and
a different mixture is observed in a detector. Such oscillations have been
observed in the neutral kaon and B--meson systems and seem now also to occur
In neutrino experiments.
 
A flavor eigenstate with a sharp momentum is a mixture of mass eigenstates with
different energies. It will oscillate in time with a well-defined oscillation
period. A flavor eigenstate with a sharp energy is a mixture of mass
eigenstates with different momenta. It will oscillate in space with a
well-defined oscillation wave length.
Many calculations describe ``gedanken" experiments which begin with states
having either a sharp momentum or a sharp energy. They require some
recipe for applying the results to a real
experiment
\cite{NeutHJL,Kayser,GoldS,Pnonexp,MMNIETO}
which is always performed with wave packets having neither sharp momenta nor
sharp energies.
 
Considerable confusion has arisen in the description of such
experiments in quantum mechanics \cite{NeutHJL,Kayser},
with questions arising about time dependence and production
reactions \cite{GoldS}, and defining precisely what
is observed in an experiment \cite{Pnonexp}.
Combining features of both the space and time oscillations can
lead to double counting.
 
This issue has been clarified\cite{GrossLip} by showing that in all oscillation
experiments the size of the neutrino source is so much smaller than the
distance between source and detector that the problem reduces to the
propagation of a linear combination of waves emitted from a point source
with well defined relative phases at the source. This wave picture uniquely
determines the relative phases at the detector, gives
all the right answers, and justifies the hand-waving used in all the
standard treatments. The particle picture is more complicated because
all momentum conservation relations must take into account the uncertainty
in the total momentum of the system resulting from the small source size,
which is orders of magnitude larger than the tiny momentum differences
between mass eigenstates.
 
\subsection{The basic quantum mechanics of flavor oscillations}
 
Treatments combining classical particle and classical wave descriptions are
often inconsistent with quantum mechanics and violate uncertainty
principles. It is inconsistent to describe a neutrino to be both a
classical point-like particle following a classical path in space-time
and also  a classical wave with a definite frequency and wave length and
a phase which is a well defined function of space-time.
The neutrino emitted in a weak interaction is a wave packet described by a
quantum-mechanical wave function, not a classical point-like particle
which travels between source and detector in a well-defined
time. The neutrino wave passes the detector during a finite time interval.
Its amplitude at the position of the detector defines the probability of 
observing the neutrino at the detector as a function of time. The flavor 
structure observed at the detector depends upon the relative phases of the mass 
eigenstate waves at the detector and upon the overlaps between them.
 
       The assumption that the mass eigenstate is simultaneously a particle
which arrives at the detector at a definite time and also a wave with a well
defined phase violates basic principles of quantum mechanics. A pulse short
enough to define a time interval exactly has no well-defined frequency and no
well-defined phase. A pulse long enough to define a phase exactly must contain
many wave lengths in space and many periods in time.
The physical neutrino in an oscillation experiment is described by a wave with
such adequate lengths in space and time. The wave defines a probability
amplitude for its observation at the detector. The exact time of detection, the
exact value of the time interval between emission and detection and the proper
time interval are therefore not predicted precisely and are given by a
probability distribution. This quantum-mechanical fluctuation in time for
the detection of a neutrino with well determined energy is just the well-known
``energy-time uncertainty relation" which makes it impossible to define
a phase and also a time interval which introduces uncertainty in energy and
frequency.
 
However, the flavor change at the detector; i.e. the change in the relative
phase of the mass eigenstates, is negligible during the time period when the
neutrino may be detected. The exact transit time of the neutrino from source to
detector is subject to unpredictable quantum-mechanical fluctuations, but the
flavor observed at the detector is well defined. Thus neutrino
oscillations can be observed in space and not in time in practical experiments
where the position of the source in space is well defined.
 
\section{Different Types of Flavor Oscillations}
 
\subsection{$K^o - \bar K^o $ Oscillations }
 
The first examples of flavor oscillations observed were in the production of
neutral kaons as flavor eigenstates $K^o$ and $\bar K^o$  propagating
in space as the nearly degenerate unstable mass eigenstates $K_L$ and $K_S$
states which decayed with long and very unequal lifetimes.
They were detected many ways - including both decays and interactions.
The mass eigenstates have very different lifetimes and are detectable by
this lifetime difference; i.e. by waiting until the $K_S$ has decayed to get
a pure $K_L$ beam. Their propagation in space as mass eigenstates $K_L$ and
$K_S$ induces oscillations between the flavor eigenstates $K^o$ and $\bar K^o$
which are observable by measurements at different points in space.
 
\subsection{$B^o - \bar B^o $ Oscillations }
 
These two nearly degenerate unstable bound states have short and very nearly
equal lifetimes. They are produced as flavor eigenstates and detected in
practice only by weak decays, where there are many decay modes.
The short lifetimes make it impossible to detect them by their strong
interactions as flavor eigenstates $B^o$ and $\bar B^o$.
Their propagation in space as mass eigenstates induces flavor oscillations
which are detected by observing their decays at different space points.
 
\subsection{Neutrino Oscillations}
 
Here we have two or three nearly degenerate stable elementary particles which
propagate without decay. They are produced and detected as flavor eigenstates.
There is no possible direct detection of the mass eigenstates. If the flavor
eigenstates are not mass eigenstates, their propagation in space as linear
combinations of mass eigenstates induces flavor oscillations.

\section{Right and Wrong Treatments of Flavor Oscillations}
 
\subsection{Common Wisdom}
 
\centerline { \bf WRONG!}
 
\centerline { $K^o$ at Rest - Propagates in Time}
 
\beq{YZY01} \ket{K^o(t)} = a(t) e^{-iE_Lt}\ket{K_L} + b(t) e^{-iE_St}\ket{K_S}
\eeq
\beq{YZY02} \langle K^o \ket{K^o(t)} = a(t) e^{-iE_Lt}\langle K^o\ket{K_L} +
b(t) e^{-iE_St}\langle K^o\ket{K_S}
\eeq
\beq{YZY03}
\langle K^o \ket{K^o(t)} = a(t) a^*(o) e^{-iE_Lt} + b(t) b^*(o) e^{-iE_St}
\eeq
 
\centerline { Probability of finding $K^o$ oscillates in time.}
 
\centerline { Oscillation frequency given by interference between}
 
\centerline { States of same momentum, different energies.}
 
\centerline { But nobody ever measures TIME!}
 
\centerline { All flavor oscillation experiments measure DISTANCES.}
 
\centerline { Oscillation wave length given by interference between}
 
\centerline { States of same energy, different momenta.}
 
\subsection{Correct Treatment}
 
\centerline { Nobody ever measures TIME!}
 
\centerline { All flavor oscillation experiments measure DISTANCES.}
 
\centerline { $K^o$ at Source - Propagates in Space }
 
\beq{YZY11}
\ket{K^o(x)} = a(x) e^{-ip_Lx}\ket{K_L} + b(x) e^{-ip_Sx}\ket{K_S}
\eeq
\beq{YZY12}
\langle K^o \ket{K^o(x)} = a(x) e^{-ip_Lx}\langle K^o\ket{K_L} +
b(x) e^{-ip_Sx}\langle K^o\ket{K_S}
\eeq
\beq{YZY13}
\langle K^o \ket{K^o(x)} = a(x) a^*(o) e^{-ip_Lx} + b(x) b^*(o) e^{-ip_Sx}
\eeq
 
\centerline { Probability of finding $K^o$ oscillates in space.}
 
\centerline { Oscillation wave length given by interference between}
 
\centerline { States of same energy, different momenta.}
 
\medskip
 
\centerline {\bf WHY SAME ENERGY?}
 
\centerline {Gives Right Answer}
 
\centerline {But how do we know it's right?}
 
\section{Paradoxes in Classical Treatments of Oscillations}
 
\subsection {Problems - Why Are States with Different Masses Coherent?}
 
\subsubsection {Energy-momentum kinematics}
 
Consider the example of a pion decay at rest into a neutrino and muon,
$\pi \rightarrow \mu \nu$. The energy $E_{\pi}$ and the momentum $p_{\pi}$
of the pion are:
\beq{YZY14}
E_{\pi} = M_{\pi}; ~ ~ ~  ~ ~ ~ p_{\pi} = 0
\eeq
where $M_{\pi}$ denotes the pion mass. Conservation of energy and momentum
then determine the energies and momenta $E_{\nu}$, $E_{\mu}$, $p_{\nu}$ and
$p_{\mu}$ of the neutrino and muon,
\beq{YZY15}
E_{\nu} = M_{\pi} - E_{\mu}; ~ ~ ~ ~ ~ ~  p_{\nu} = -p_{\mu}
\eeq
 
The mass of the neutrino $M_{\nu}$ is then determined by the relation
 
\beq{YZY16}
M_{\nu}^2  = (M_{\pi} - E_{\mu})^2 - p_{\mu}^2
\eeq
 
This is just a ``Missing Mass" experiment. The value of $M_\nu$ is uniquely
determined and there can be no interference between states of different mass.
 
\subsubsection {Space-time measurements}
 
Consider a neutrino created at the space-time point $(x=0, ~ t=0)$ with
momentum $p$. It is detected at the position of a detector, $(x=x_d)$.
The time of detection, $t_d = x_d/v$ depends upon the velocity of the neutrino.
It the neutrino is a linear combination of two mass eigenstates with masses
$m_1$  and $m_2$, they will have different velocities,
 
\beq{YZY20}
v_1 = {{p}\over{m_1}}; ~ ~ ~ v_2 = {{p}\over{m_2}}
\eeq
 
They will therefore arrive at the detector with different arrival times,
\beq{YZY21}
t_1 = {{x_d\cdot m_1}\over {p}}~ ~ ~ t_2 = {{x_d\cdot m_2}\over {p}}
\eeq
 
The detector will therefore detect either one or the other. There will be no
coherence between mass eigenstates, no interference and no oscillations.
 
\subsection{Solutions - Wave-particle duality provides coherence}
 
\subsubsection{Common Feature of all Flavor Oscillation Experiments}
 
The flavor-oscillating particle is produced as a flavor eigenstate by a
localized source in space. It is detected at a large distance $(x_d)$ compared
to the source size $(x_s)$. If the flavor eigenstate is produced with a sharp
energy and is a linear combination of mass eigenstates with masses $m_1$ and
$m_2$, they have momenta $p_1$ and  $p_2$. Space oscillations arise from
interference between $p_1$ and  $p_2$.
 
The uncertainty principle requires a momentum uncertainty in the particle
wave-packet
$\delta p_W \approx \hbar /x_s$.
This will also produce an uncertainty in the energy.
Coherence between mass eigenstate waves will occur if the momentum difference
between the different mass eigenstates $with$ $the$ $same$ $energy$,
$|p_1 - p_2|_E$ is much smaller than momentum uncertainty in the wave packet
$|p_1 - p_2|_E << \delta p_W$ and give rise to spatial oscillations.
 
\subsubsection{Lipkin's Principle - If you can measure it you can measure it! }
 
\centerline {\bf PROOF }
 
Any sensible experiment must have an oscillation wave length $\lambda$ much
larger than source size.
\beq{WW701}
\lambda \approx {{\hbar} \over {|p_1 -p_2|_E}}  >> x_s
\eeq
The momentum uncertainty must then be much larger than the momentum difference
between the mass eigenstates.
\beq{WW702}
\delta p_W \approx {{\hbar }\over {x_s}}
>>
{{\hbar} \over {\lambda}} \approx |p_1 -p_2|_E
\eeq
Thus any sensible experiment will have $p_1-p_2$ coherence.
 
Note that this implies that the initial state of any realistic flavor
oscillation experiment does not have a sharp four-momentum. The
quantum-mechanical fluctuations in this four-momentum
required by the uncertainty principle are always much larger than the
four-momentum differences between the different mass eigenstates which produce
oscillations. They are therefore also much larger than any four-momentum
differences between the states of other particles recoiling against these mass
eigenstates. Thus any possible effects like induced oscillations which use
four-momentum conservation to obtain a precise knowledge of the recoil momentum
are destroyed by these quantum-mechanical four-momentum fluctuations.
 
\section{Right and Wrong Ways to Treat Flavor Oscillations}
 
\subsection {THE RIGHT WAY}
 
\subsubsection {The Problem}
 
A particle with definite flavor is created at a source. This particle is a
linear combination of mass eigenstate waves with amplitudes and phases
determined by the mixing dynamics. The mass eigenstates propagate independently
with no interactions (we exclude the MSW interactions for the present) in a
manner described by the Schroedinger or Dirac equation.
The relative phases of different mass eigenstate waves  change during
propagation in space.
 
The problem is to calculate the flavor of the particle measured at a remote
detector which depends upon the relative phases of the mass eigenstates at that
point.
 
\subsubsection {The Solution}
\begin {enumerate}
 
\item{Solve the free Schroedinger or Dirac Equation. This solution is trivial
with no need for fancy field theory or Feynman diagrams. The presence of
mixtures of noninteracting mass states provide no problem.}
 
\item {Introduce the proper initial conditions at the source. This means
defining a wave packet whose behavior in space and time describe the real
experiment.}
 
\item {Get the answer for what is observed at the detector by evaluating the
solution of the propagation equations at the detector.}
 
\end {enumerate}
 
 \subsubsection {The Question}
 
WHY DOESN'T ANYONE DO THIS?
 
\subsection {WHAT EVERYONE DOES INSTEAD - HAND WAVING!}
 
\begin {enumerate}
 
\item {Solve the wrong problem - Flavor oscillations in time. Nobody measures
oscillations in time.}
\item {Obtain a correct but useless irrelevant answer - the frequency of
oscillations in time.}
\item {Handwave to convert the irrelevant answer to the wrong problem into the
answer to the right problem; to convert the frequency of oscillations in time
to the wave length of oscillations in space.}
\item {Right hand waving by using $x = v t$ and choosing the right value for
$v$ gives the right answer.}
\item {Wrong hand waving gives the wrong answer.}
\item {All results in textbooks and in papers used by experimenters and
phenomenologists to analyze data have used the right hand waving and get the
right answer}
\item {The literature is still flooded with papers using the wrong hand
waving, publishing wrong answers, and confusing many people.}
\end {enumerate}
 
\section {Real \& Gedanken $\nu$-oscillation Experiments}
 
A mixture of two or more mass eigenstates is created by a source and a
different
mixture is observed in a detector. If the initial state is a flavor eigenstate
with a sharp momentum the mass eigenstates have  different energies and
oscillations in time are observed with a  well-defined oscillation period.
If the initial state is a flavor eigenstate with a sharp energy, the mass
eigenstates have different momenta and oscillations in space are observed with
a well-defined oscillation wave length. Experiments always measure oscillations
in space; whereas conventional wisdom describes oscillations in time.
 
We now show in a simple example how the description of a
time-dependent non-experiment can lead to ambiguities and confusion.
Consider neutrino oscillations in one dimension with two mass eigenstates.
We assume a $45^o$ mixing angle for convenience so that the states
$\ket{\nu_e}$ and $\ket{\nu_\mu}$ are equal mixtures with opposite
relative phase of the mass eigenstates denoted by
$\ket {\nu_1}$ and
$\ket{\nu_2}$ with masses denoted respectively by $m_1$ and $m_2$.
\beq{YZY31}
\ket{\nu_e} = (1/\sqrt 2)(\ket {\nu_1}+ \ket{\nu_2}); ~ ~ ~ ~
\ket{\nu_\mu} = (1/\sqrt 2)(\ket {\nu_1} - \ket{\nu_2})
\eeq

\subsection {The Gedanken Time Experiment}
 
Consider the ``non-experiment" often described in which a
a $\nu_e$ is produced at time t=0 in a state of
definite momentum $p$. The energies of the
$\nu_1$ and $\nu_2$ components
denoted by $E_1$ and $E_2$ will be different and given by
\beq{YZY41}E_1^2 = p^2 + m_1^2; ~ ~ ~ ~ ~ ~ E_2^2 = p^2 + m_2^2  \eeq
 
Let $\ket{\nu_e(t)}$ denote this linear combination of $\ket {\nu_1}$ and
$\ket{\nu_2}$ with energies $E_1$ and $E_2$ which is a pure $\ket{\nu_e}$
at $t=0$. The $\ket{\nu_e}$ and $\ket{\nu_\mu}$ components of this wave
function will oscillate as a function of $t$ in a manner described by
the expression
\beq{YZY42} \left |{{ \langle \nu_\mu\ket{\nu_e(t)}}\over
{ \langle \nu_e\ket{\nu_e(t)}}}\right |^2=
\left |{{e^{iE_1 t} - e^{iE_2 t}}\over {e^{iE_1 t} + e^{iE_2 t}}}\right |^2=
\tan^2 \left({{(E_1 - E_2)t}\over{2}}\right) =
\tan^2 \left({{(m_1^2 - m_2^2)t}\over{2(E_1 + E_2)}}\right)
\eeq
 
This is a ``non-experiment" or ``gedanken experiment". To compare this result 
with a real experiment which measures space oscillations the gedanken time 
dependence must be converted into a real space dependence. Here troubles and 
ambiguities arise and the need for hand-waving.
 
\subsubsection {Handwaving - Method A}
 
One can simply convert time
into distance by using the relation
\beq{YZY51} x = vt = {{p}\over{E}} \cdot t                      \eeq
where $v$ denotes the velocity of the $\nu$ meson. This immediately gives
\beq{YZY52} \left |{{ \langle \nu_\mu\ket{\nu_e(t)}}\over
{ \langle \nu_e\ket{\nu_e(t)}}}\right |^2=
\tan^2 \left({{(m_1^2 - m_2^2)t}\over{2(E_1 + E_2)}}\right) \approx
\tan^2 \left({{(m_1^2 - m_2^2)x}\over{4p}}\right)
\eeq
where the
small differences between $p_1$ and $p_2$ and between $E_1$ and $E_2$
are neglected.
 
\subsubsection {Handwaving - Method B}
 
However, one can also argue that the $\nu_1$ and $\nu_2$ states with the
same momentum and different energies also have different velocities,
denoted by $v_1$ and $v_2$ and that they therefore arrive at the point
x at different times $t_1$ and $t_2$,
\beq{YZY53} x = v_1t_1 = {{p}\over{E_1}}\cdot t_1
= v_2t_2 = {{p}\over{E_2}}\cdot t_2                    \eeq
One can then argue that the correct interpretation of the
time-dependent relation for measurements as a function of $x$ is
\beq{YZY54} \left |{{ \langle \nu_\mu\ket{\nu_e(x)}}\over
{ \langle \nu_e\ket{\nu_e(x)}}}\right |^2=
\left |{{e^{iE_1 t_1} - e^{iE_2 t_2}}\over {e^{iE_1 t_1} +
e^{iE_2 t_1}}}\right |^2=
\tan^2 \left({{(E_1t_1 - E_2t_2)}\over{2}}\right) =
\tan^2 \left({{(m_1^2 - m_2^2)x}\over{2p}}\right)
\eeq
This differs from the relation (\ref{YZY52}) by a factor of 2 in the
oscillation wave length. If one does not consider directly the result of a
real experiment but only the two different interpretations of the
gedanken experiment, it is not obvious which is correct. Questions also arise 
regarding the use of phase velocity or group velocity in eqs.(\ref{YZY52}) and 
(\ref{YZY54})
 
\subsection {The real experiment - measurement directly in space}
 
All this confusion is
avoided by the direct analysis of use of the result of the real experiment.
In an experiment where a $\nu_e$ is produced at x=0 in a state of
definite energy $E$, the momenta of the $\nu_1$ and $\nu_2$ components
denoted by $p_1$ and $p_2$ will be different and given by
\beq{YZY55}p_1^2 = E^2 - m_1^2; ~ ~ ~ ~ ~ ~ p_2^2 = E^2 - m_2^2  \eeq
Let $\ket{\nu_e(x)}$ denote this linear combination of $\ket {\nu_1}$ and
$\ket{\nu_2}$ with momenta $p_1$ and $p_2$ which is a pure $\ket{\nu_e}$
at $x=0$. The $\ket{\nu_e}$ and $\ket{\nu_\mu}$ components of this wave
function will oscillate as a function of $x$ in a manner described by
the expression
\beq{YZY56}
\left |
{{ \langle \nu_\mu\ket{\nu_e(x)}}\over { \langle \nu_e\ket{\nu_e(x)}}}
\right |^2
=
\left |
{{e^{ip_1 x} - e^{ip_2 x}}\over {e^{ip_1 x} + e^{ip_2 x}}}
\right |^2
=
\tan^2 \left({{(p_1 - p_2)x}\over{2}}\right) \approx
\tan^2 \left({{(m_1^2 - m_2^2)x}\over{4p}}\right)
\eeq
These are just the normal neutrino oscillations, and the results agree with
those (\ref{YZY52}) obtained by handwaving A.
 
We immediately note the analogous implications for all
experiments measuring flavor oscillations.
Calculations for neutrino oscillations
in time describe non-experiments. Times are never measured in the
laboratory; distances are measured. When correlated decays of two mesons
will be measured in an asymmetric B factory, the points in space where
the two decays will be measured in the laboratory, not the time
difference which appears in many calculations.
 
When a $\nu_e$ is produced at x=0 with energy $E$, its mass eigenstates
propagate in space and their relative phase changes produce
$\ket{\nu_e}$ and $\ket{\nu_\mu}$ oscillations in space. The simple argument
using handwaving A is right. The treatment is completely relativistic and
needs no discussion of time dependence or ``proper times".
 
\centerline {\bf But is the use of a sharp energy really correct?}
 
\subsection {Another Approach with Different $E$ $and$ Different $p$}
 
The interference has also been considered \cite{GoldS} between two states
having both different $E$ $and$ different $p$ produced at the point $x=0,t=0$..
\beq{YZY57} \left |{{ \langle \nu_\mu\ket{\nu_e(x,t)}}\over
{ \langle \nu_e\ket{\nu_e(x,t)}}}\right |=
\left |{{e^{i(E_1t-p_1x)} - e^{i(E_2t-p_2x)}}\over {e^{i(E_1t-p_1x)} +
e^{i(E_2t-p_2x)}}} \right |=
\tan \left({{(E_1 - E_2)t- (p_1 - p_2)x}\over{2}}\right)
\eeq
We now find that we can get the same result as the above treatment with a sharp
energy (\ref{YZY56}) if we choose the time that the wave appears at the detector
as the time after traveling with the
mean group velocity $\langle v_{gr} \rangle$,
\beq{YZY58}t = {{x}\over{\langle v_{gr} \rangle}} = x\cdot
{{E_1+E_2}\over{p_1+p_2}}
\eeq
\beq{YZY59} \left |{{ \langle \nu_\mu\ket{\nu_e(x)}}\over
{ \langle \nu_e\ket{\nu_e(x)}}}\right |=
\tan \left({{[(E_1^2 - E_2^2)- (p_1^2 - p_2^2)]x}\over{2(p_1+p_2)}}\right) =
\tan \left({{(m_1^2 - m_2^2)}\over{2(p_1+p_2)}}\right)\cdot x
\eeq
This result is simply interpreted in the wave picture. Eq. (\ref{YZY57})
holds at all points in space and time, and is due to the difference in the
$phase$ velocities of the two mass eigenstate waves. To apply this to the 
detector, we substitute the position of the detector and the time at which the 
neutrino is detected. There is only a single time, not two times as in 
eq.(\ref{YZY54}) obtained by Handwaving B. Although the centers of the wave 
packets move apart, the neutrino is detected for both wave packets at the same 
single time. 
 
However, one can question the use of the expression value (\ref{YZY58})
determined by the mean group velocity. Since the wave
packets pass the detector during a finite time interval, the detection time
$t$ to be substituted into eq. (\ref{YZY57})  can be $any$ $time$ during
which the wave amplitude is finite at the detector. There is therefore a
spread $\delta t$ in the detection time which will give rise to a spread in
the relative phase $\delta \phi$ between the two mass eigenstates.
\beq{YZY60} \delta \phi ={{(E_1 - E_2)}\over{2}}\cdot \delta t
\approx {{(E_1 - E_2)}\over{2 \delta E}}
\eeq
where $\delta E = 1/\delta t$ is the spread in energy required by the
uncertainty principle for a wave packet restricted in time to an
interval $\delta t$. We thus see that the uncertainty $\delta \phi$ will
be of order unity and wash out all oscillations unless the energy
difference $E_1 - E_2$ between the two interfering mass eigenstates
is much smaller than the energy spread in the wave packet. We are
therefore reduced to case described by eq. (\ref{YZY56}) and the necessity
for use of a sharp energy to reneder oscillations observable.
 
The use of  sharp energies has been justified\cite{GrossLip,Leo} and
is discussed in detail below. First we review carefully what is known
in a realistic neutrino oscillation experiment and what cannot be
known because of quantum mechanics and the uncertainty principle.
 
\section{What do we know about Flavor Oscillations}
 
\subsection{A General Guide to knowledge}
 
\centerline {\bf My Father Used to Tell Me}
 
\centerline {``If you would know what you don't know,}
 
\centerline {You would know more than you know"}
 
\centerline {**********************************************}
 
\centerline {\bf Quantum Mechanics Tells Us}
 
\centerline {You can't know everything}
 
\centerline {If you know the position of a neutrino source,
you don't know its momentum}
 
\centerline {**********************************************}
 
\centerline {\bf Guide to Flavor Oscillations}
 
\centerline {Use what you can know}
 
\centerline {Don't cheat by pretending you know what you can't know}
 
\centerline {**********************************************}
 
\centerline {\bf Examples of What We Can't Know}
 
\centerline {The total momentum of a neutrino source in any experiment}
 
\centerline {The momentum of muon, $\Lambda$ or other particle recoiling
against a mass eigenstate}
 
\centerline {Exact center-of-mass system for fixed target experiment}
 
\centerline {Neutrino transit time from source to detector}
 
\centerline {All these are smeared by the uncertainty principle}
 
\subsection{What do we really know and really not know?}
 
\centerline {We know there is a neutrino source}
 
\centerline {We know the position of the source}
 
\centerline {We know the flavor of the neutrino emitted by the source}
 
\centerline {We do not know the time of emission!}
 
\centerline {We do not know the momentum of the source}
 
\centerline {**********************************************}
 
\centerline {We know there is a neutrino detector}
 
\centerline {We know the position of the detector}
 
\centerline {We know the sensititivity of the detector to neutrino flavor}
 
\centerline {We do not know the time of detection!}
 
\centerline {\bf All books cheat by pretending we know what you can't know}
 
\centerline {**********************************************}
 
\subsection{RECOIL is a RED HERRING! RECOILS are unobservable}
 
\centerline{Recoil momenta of muons, $\Lambda$'s etc. given only
by proability distributions}
 
\centerline{Oscillations of recoil particles completely washed out by
quantum-mechanical fluctuations}
 
\centerline {**********************************************}
 
\subsection{TIME is a RED HERRING! Nobody measures TIME!}
 
\centerline{Solar Neutrino Experiments}
 
\centerline{Atmospheric Neutrino Experiments}
 
\centerline{Reactor Neutrino Experiments}
 
\centerline{Accelerator Neutrino Experiments}
 
\centerline {\bf None of them measure TIME!}
 
\centerline {\bf Nobody wants to measure TIME!}
 
\centerline {\bf Nobody would know what to do with a TIME measurement!}
 
\section{The Kinematics of Fixed Target Experiments}
 
The complete description of a flavor oscillation experiment requires knowledge
of the density matrix for the flavor-mixed state. This depends upon the
production mechanism and possible entanglements with other degrees of freedom
as well as on other dynamical factors which are often ignored.
 
One example of such a generally ignored dynamical factor is the force on a
proton in a fixed-target experiment. This proton is not free. To keep it in
a solid target it must be constrained by some kind of effective
potential with characteristic lattice energies like Debye temperatures.
This energy scale is of the order of tens of millivolts and not at all
negligible in comparison with mass differences between flavor eigenstates.
In a simple potential model the proton is initially
in some energy level with a well defined total energy. But there are large
variations in its potential and kinetic energies. Thus the kinetic energy and
momentum of the proton are not sharply defined. The
bound proton is not strictly on shell and arguments of Galilean and Lorentz
invariance and separation of center-of-mass motion may not hold for the
kinematics of the production process if the degrees of freedom producing the
binding are neglected.
 
Consider for example the reaction
\beq{YY101}
\pi^- + p \rightarrow K^o + \Lambda
\eeq
If the energies and momenta of the pion beam, the target proton, and the
outgoing $\Lambda$ are known, the energy, momentum and mass of the
outgoing kaon are determined by energy and momentum conservation. If,
however, the energy and momentum of the target proton differ by small
amounts
$\delta E$ and $\delta \vec p$ from the values for a free proton
at rest, the squared mass of the kaon determined from conservation laws
is given to first order in the small quantity
$\delta \vec p$ by
\beq{YY102}
M_K^2  = M_K(o)^2 +
\delta M_K^2 ~ ~ ~ ; ~ ~ ~ \delta M_K^2  \approx
- 2 \delta \vec p \cdot (\vec p_\pi - \vec p_\Lambda)
\eeq
where $M_K(o)$ denotes the value of the kaon mass that is obtained from
the conservation laws when $\delta E$ and $\delta \vec p$ are neglected
and we note that
$\delta E$ is of second order in $\delta \vec p$ and can be neglected
to this approximation. Let us assume that the target proton is bound in
a solid with a characteristic frequency $\omega$; e.g. the Debye or
Einstein temperature of a crystal. This then sets the scale of the
kinetic energy of the bound proton. Thus
\beq{YY103}
  |\delta \vec p| = O(\sqrt{M_p\cdot \omega})
~ ~ ~ ; ~ ~ ~ \delta M_K^2  = O(\sqrt{M_p\cdot \omega})
\cdot |\vec p_\pi - \vec p_\Lambda|
\eeq
Since $\omega$ is of order $10^{-2}$ ev., while
$M_K$, $M_p$, $\vec p_\pi$ and $\vec p_\Lambda$ are all of order 1 GeV,
we see that $|\delta \vec p|$ and $\delta M_K$ are
of order 3 KeV. This is so much larger than
the mass difference $ 3 \times 10^{-6}$ ev. that any discussion of
detecting recoil effects due the kaon mass difference is simply
ridiculous. Since the momentum of the center of mass in this experiment
has an uncertainty of 3 KeV due to the continuous exchange of momentum
between the target proton and the forces binding it to the target, one
cannot define a center-of-mass system for the beam and proton and ignore
the rest of the target. Galilean and Lorentz transformations are clearly
not valid at the scale of the kaon mass difference, without also
transforming the macroscopic target to the moving frame.
 
In the language of the parton model the target proton might be considered as a
parton moving in a sea of ``brown muck".
Measurements of energy and momentum of incoming and outgoing
particles then determine the energy and momentum distribution of the
``parton" proton in the initial state. However, this does not work for the
same reason that the parton model cannot describe the photoelectric effect
in which an electron is ejected from an inner shell by the absorption of a
photon. One must understand the dynamics of the binding and know the bound
state wave function and the ionization energy to predict the results of a
photoelectric experiment. Knowing the momentum distribution of the electron
``parton" is not enough. Similarly describing the finite momentum spread of a
target proton by a momentum distribution is not enough to enable prediction of
the results of an experiment using the reaction (\ref{YY101}) to the accuracy
required for the determination of the kaon mass difference. One
must know a wave function or density matrix as well as an ionization or
dissociation energy in order to take subtle coherence effects and energy
conservation into account.
 
If however, one is only interested in determining the kaon mass difference and
not in the precise measurements of recoil momenta on that scale, a detailed
knowledge of the bound state wave function is not necessary. One only needs to
know that the bound state wave function in momentum space is sufficiently wide
to produce full coherence between components of the same energy with different
mass and different momenta. The measured oscillation wave length then
determines the mass difference to the same precision with which the wave length
is determined. There is no need to measure momenta at the kilovolt level. This
is shown in detail below.
 
The required coherence is between states of the same energy and
different momenta, rather than vice versa.
That energy and momentum conservation are not on
the same footing is seen here as the same physics that describes the
photoelectric effect and describes bouncing
a ball elastically against the earth with energy conservation and no
momentum conservation. In each case the relevant degrees of freedom are
in interaction with a very large system which can recoil with arbitrary
momentum and negligible kinetic energy.
 
\section{What Is Measured in Real Neutrino Oscillation Experiments}
 
\subsection{A single mass state passes a detector}
 
\centerline {NEUTRINO INCIDENT ON DETECTOR IS A WAVE!}
 
\centerline {Has finite length - passes detector in finite time interval}
 
\centerline {Square of amplitude at time $t$ gives probability of detection}
 
\centerline {DETECTION TIME WITHIN WAVE PACKET UNPREDICTABLE!}
 
\centerline {Time of detection generally not measured}
 
\centerline {Precise time measurement gives no useful information!}
 
\subsection{Two overlapping mass states pass detector}
 
\centerline {NEW INGREDIENT: Neutrino flavor depends on relative phase}
 
\centerline {Still finite length - finite time interval}
 
\centerline {Square of amplitude at time $t$ gives probability of detection}
 
\centerline {DETECTION TIME WITHIN WAVE PACKET UNPREDICTABLE!}
 
\centerline {Relative phase changes with space and time in packet}
 
\centerline {Negligible phase change with time at fixed detector!}
 
\centerline {DETECTION TIME WITHIN WAVE PACKET STILL USELESS!}
 
\section{An optical guide to neutrino oscillations}
 
\subsection{A Faraday-rotated optical beam}
 
As an  instructive electromagnetic analog to quantum mechanical particle
flavor oscillations consider the propagation of a Faraday-rotated polarized
optical beam. We examine the case where a source emits vertically polarized
light through a medium in which a magnetic field produces Faraday rotations.
The parameters are chosen so that the plane of polarization is rotated
by $90^o$ between the source and detector. The light then reaches the detector
horizontally polarized. Because of the presence of the medium, the light
travels with phase and group velocities which are different from $c$. The
states of right and left handed circular polarization are analogous to the
neutrino mass eigenstates, which propagate unchanged through space. The states
of plane polarization are analogous to neutrino flavor eigenstates which
undergo oscillations while propagating in space. In this picture one can
consider neutrino flavor as an intrinsic degree of freedom described by $SU(n)$
rotations in an abstract space where $n$ is the number of flavors.
 
\subsubsection { A classical wave picture }
 
In a classical wave picture the light is a coherent linear combination of
left-handed and right handed circularly polarized light beams which travel
with slightly different velocities. The tiny velocity difference produces a
change in the relative phase of left-handed and right handed components and
rotates the plane of polarization.
 
\subsubsection { Quantum photon picture }
 
But light is quantized and consists of photons. What happens to a single
vertically-polarized photon? Will it arrive horizontally polarized at the
detector? The left-handed and right-handed components have different velocities
and will arrive at the detector at different times.
 
This is a standard quantum-mechanical problem occurring whenever a beam of
polarized particles passes through a field which would classically rotate the
direction of polarization. Sometimes the components remain coherent and rotate
the polarization. Sometimes they split to produce a Stern-Gerlach experiment.
 
\subsubsection {Back to classical wave picture}
 
For more intuition upon when there is coherence and when there is Stern-Gerlach
we consider a classical source emitting classical pulses of finite length. They
are therefore not monochromatic; there is a chromatic aberration that fuzzes
the polarization. There is a classical uncertainty principle known to every
electronic engineer. To define the time of a short pulse to a precision
$\delta t$ one needs a finite band width $\delta \nu$ which satisfies  the
classical uncertainty principle $\delta \nu \cdot \delta t \approx O(1)$.
 
The two pulses with left and right circular polarization have different
velocities and gradually move apart. During the separation period there is
a coherent overlap region with plane polarization and incoherent forward and
backward zones with opposite circular polarizations.
 
\subsubsection {Back to quantum photon picture }
 
We now can quantize this picture and see that a photon can be detected either
in the overlap region or in the forward or backward zones. A photon produced
in the overlap region is horizontally polarized; a photon produced in the
forward or backward zones is circularly polarized. The amplitude at the
detector at time $t$ gives probability of detecting a photon at time $t$.
For quantized waves, Planck introduces $E=h\nu$ to get the quantized uncertainty
relation $\delta E \cdot \delta t \approx O(h)$. But the uncertainty between
frequency and time and between position and wave-length are already there in the
wave picture. It is the quantum-mechanical wave-particle duality that makes
these into uncertainties between energy and time and between position and
momentum.
 
\subsection{A Faraday-Rotated Polarized Radar Pulse}
 
To get a quantitative picture let us consider the propagation of a plane
polarized microwave radar pulse through a medium containing a magnetic field
in which Faraday rotations occur. Let the difference in velocities between the
left-handed and right-handed polarization states be tiny, of order one part
per million,
\beq{ZZ301}
  {{\delta v}\over{v}} = 10^{-6}
\eeq
This velocity difference introduces a relative phase shift between the two
circularly polarized waves observed as a rotation of the plane of polarization
between the transmitter and receiver.
We first consider the classical wave picture and then introduce the
quantum particle picture by considering individual photons.
 
Consider a pulse of one microsecond duration and a wave length of one
centimeter traveling at very near the velocity of light. We assume that
the deviations in velocity produced by the medium and magnetic field are
smaller than one part per million and negligible for rough estimates.
The length of the wave train or wave packet in space $L_w$ is
\beq{ZZ302}
L_w = 3\times 10^{10}\cdot 10^{-6} = 10^4 ~ \rm cm.
\eeq
Both
the size of the transmitter and the size of the receiver are small relative
to the length of the wave train, which contains $10^4$ wave lengths. The
frequency of the microwave radiation is seen to be
\beq{ZZ303}
\nu = 3\times 10^{10}
~ ~ \rm cycles
\eeq
or 30,000 megacycles. However, the radiation is not monochromatic. The
frequency spectrum of a one microsecond pulse must have a finite band width
of the order of one megacycle.
 \beq{ZZ304}
\delta \nu \approx 10^{6}
~ ~ {\rm cycles}
= \nu/3,000
\eeq
 
Since the velocities of the right-handed and left-handed pulses are different,
the two wave packets eventually separate. If the receiver is sufficiently
distant, it receives two one-microsecond pulses circularly polarized in
opposite directions. We examine the interesting domain when the
the distance between transmitter and receiver is sufficiently small so that
the overlap between the two circularly polarized wave packets is essentially
$100\%$; e.g. if the centers of the wave packets have separated by 10 cm. which
is negligible compared to the 100 meter lengths of the packets but sufficiently
large so that the plane of polarization has undergone 10 complete Faraday
rotations between the transmitter and receiver. If polarization measurements
are made between the transmitter and receiver, 10 oscillations will be observed
over this distance.
 
Since  $\delta v/ v = 10^{-6}$, ten
oscillations will be observed after the waves have traversed a distance of
ten million wave lengths; i.e. 100 kilometers. The oscillation wave length
will be ten kilometers. The transit time of the wave will be
 
 \beq{ZZ305}
\delta t = {{10^7}\over{3\cdot 10^{10} }} = (1/3)\cdot 10^{-3}~ ~
{\rm sec.}
\eeq
or (1/3) millisecond.
 
The description in quantum mechanics is seen by examining the case
where the transmitter is sufficiently weak and the receiver sufficiently
sensitive so that individual photons can be counted in the receiver. The one
microsecond pulse observed at the detector is seen as individual photons
whose time of arrival at the detector are equally distributed over the one
microsecond interval. There is thus a fluctuation of one microsecond
in the times of arrival of an individual photon. This gives an uncertainty
in the transit time of 3 parts per thousand. In any calculation of
the velocity of the photon from the measured time of arrival
after it traversing a distance of 100 kilometers, the uncertainty of the
arrival time produces an uncertainty in the velocity of 3 parts per thousand.
This is enormous compared to the resolution of one part per million required
to distinguish between the velocities of the two circularly polarized
components. In principle one could measure the velocity difference by
measuring the centroid of the arrival time distribution with sufficient
precision. In practice this is out of the question.
 
The photons arriving at the receiver remain coherent mixtures of the two
circularly polarized states. The polarization observed at the detector is just
the polarization defined by the classical Faraday rotation; i.e. the relative
phase of the two circularly
polarized components arising from their traveling at different velocities.
The exact time of arrival of an individual photon plays no role here. The
quantum-mechanical uncertainty in the time arrival arising from the finite
time duration of the pulse makes it impossible to determine the velocity of
the photon to the precision needed to distinguish between the velocities of
the two circularly polarized components.
 
If the detector is 10,000 kilometers or $10^9$ cm. from the source, the centers
of the two waves will have separated by $10^3$ cm or $(1/10)L_w$. The
probability for observing a photon will now have spread to an interval of
1.1 microsecond, The photons detected in the central 0.9 microseconds of this
interval will still have the polarization defined by the classical Faraday
rotation. The first first and last intervals of 0.1 microseconds will now be
left-handed and right-handed circularly polarized.
As the distance is increased, the circularly polarized leading and trailing
edges of the wave becomes greater until the wave separates into two
one-microsecond pulses circularly polarized in opposite directions.          .
 
The essential feature of this description is the necessity to create a wave
train which contains a large number of cycles. This allows the different
components of the wave packet traveling with different velocities to separate
by a small number of cycles without appreciably affecting the overlap between
these components. This is also the essential feature of any flavor oscillation
experiment where a source creates a wave packet containing a sufficiently large
number of cycles so that displacements of a few cycles between the packets of
different mass eigenstates traveling with different velocities produce a
relative phase shift at the detector of the order of one cycle without
appreciably affecting the overlap between the wave packets. Exact measurements
of transit times between source and detector play no role, as they are subject
to quantum-mechanical fluctuations arising from the condition that the length
of the wave packet must contain a sufficient number of cycles to enable the
definition of a phase and a frequency.
 
The above optical analog is easily taken over into the description of
particle flavor oscillations.  The flavor eigenstates are
analogous to spin polarization eigenstates, and the neutrino oscillations
are describable as rotations in some abstract flavor-spin space. The fact
that all experiments in which oscillations can be measured involve sources
which are very small in comparison with the oscillation wave length enable
a description in which waves are emitted from a point source with a
definite polarization state in this flavor-spin state.
 
\section{A Universal Boundary Condition Approach}
 
\subsection {Resolution of Confusion}
 
We have noted that the proper solution for the flavor oscillation problem is
simply to solve the free Schroedinger or Dirac equation and introduce the
proper initial conditions at the source. The reason why nobody ever does this
is because the initial conditions at the source are generally very
complicated and not known. This is the reason for the general procedure
of solving gedanken experiments and hand waving.
 
However, it has now been shown\cite{GrossLip,Leo} that it is not necessary to
know all details of the initial conditions in order to obtain the desired
results. Much confusion has been resolved\cite{GrossLip} by noting
and applying one simple general feature of all practical experiments.
The size of the source is small in comparison with the
oscillation wave length to be measured, and
a unique well--defined flavor mixture is emitted by the source; e.g. a $\nu_e$
in a $\nu$ oscillation experiment. The particles emitted from the source must
leave the source before their flavor begins to oscillate. They are
therefore described by a wave packet which satisfies a simple general
boundary condition: the probability amplitude for finding a particle having
the wrong flavor; e.g. a $\nu_\mu$ at the source must vanish for all times.
There should be no flavor oscillations at the source.
 
This boundary condition requires factorization of the flavor and time dependence
at the position of the source. Since the energy dependence is the Fourier
transform of the time dependence, this factorization also
implies that the flavor dependence of the wave packet is independent of energy
at the position of the source.
In a realistic oscillation experiment the relative phase is important when the
oscillation length is of the same order as the distance between the
source and the detector.
In that case this
flavor--energy factorization holds over the entire distance between the source
and detector. The boundary condition then determines
the relative phase of components in the wave function with
different mass having the same energy and different momenta. Thus
any flavor oscillations observed as a function of the distance between the
source and the detector are
described by considering only the interference between
a given set of states having the same energy. All questions of coherence,
relative phases of components in the wave function with different energies and
possible entanglements with other degrees of freedom are thus avoided.
 
Many formulations describe flavor oscillations in time produced by
interference between states with equal momenta and different energies.
These ``gedanken" experiments
have flavor oscillations in time over all space including the source.
The ratio of the wave length of the real spatial oscillation to the period of
the gedanken time oscillation has been shown\cite{GrossLip} to be given by
the group velocity of the wave packet.
 
\subsection {Explicit Solution of Oscillation Problem}
 
We now present a rigorous quantitative treatment of the above argument and
show how the results of a flavor oscillation experiment can be
predicted without solving all the problems of production, time behavior
and coherence. If oscillations are observable, the dimensions of the source
must be sufficiently small in comparison with the distance to the detector
and the oscillation wave length to be measured so that the particle leaves
the source with its original flavor. The distance traversed by the particle
in leaving the source is too small in comparison with the oscillation wave
length for any significant flavor change to occur. It is therefore a good
approximation to consider the outgoing wave to be produced by a
point source at the origin. The wave length in space
of the oscillation can then be shown to be completely determined by the
propagation dynamics
of the outgoing wave in space and the boundary condition that the
probability of observing a particle of the wrong flavor at the position of the
source at any time must vanish for all times. Note that the exact time in
which the particle is produced is not necessarily determined. The wave packet
describing the particle must generally
have a finite spread in time at the source
position. But whenever it is produced in time, it leaves the source in space
still with its original flavor.
 
We choose
for example a neutrino oscillation experiment with a source of
electron neutrinos.
The neutrino wave function for this experiment may be a very
complicated wave packet, but a sufficient condition for our analysis is to
require it to describe a pure
$\nu_e$ source at $x = 0$; i.e. the probability of finding a
$\nu_\mu$ or $\nu_\tau$ at $x = 0$ is zero.
 
This boundary condition requires factorization of the flavor and time
dependence
at the position of the source. Since the energy dependence is the Fourier
transform of the time dependence, this factorization also
implies that the flavor dependence of the wave packet is independent of energy
at the position of the source.
 
We write the neutrino wave function as an expansion in energy eigenstates
satisfying the condition that it must
avoid spurious flavor oscillations at the source and therefore be
a pure $\nu_e$ state at $\vec x = 0 $ for a finite length of time.
\beq{WW1a}
\psi = \int g(E) dE e^{-iEt}\cdot \sum_{i=1}^3 c_i e^{i p_i\cdot x}
\ket {\nu_i}~ ~ ~ ; ~ ~ ~
\sum_{i=1}^3 c_i \bra {\nu_i}\nu_\mu \rangle =
\sum_{i=1}^3 c_i \bra {\nu_i}\nu_\tau \rangle = 0
\eeq
where $\ket {\nu_i}$ denote the three neutrino mass eigenstates and the
coefficients $c_i$ are energy-independent.
The momentum of each of the
three components is determined by the energy and the neutrino masses. The
propagation of this energy eigenstate, the relative phases of its three mass
components  and its flavor mixture at the detector are completely determined by
the energy-momentum kinematics for the three mass eigenstates.
 
The flavor mixture at the detector given by substituting the
detector coordinate into Eq. (\ref{WW1a})
can be shown to be the same for all the
energy eigenstates except for completely negligible small differences.
For example, for the case of two neutrinos with energy $E$ and mass eigenstates
$m_1$ and $m_2$ the relative phase of the two neutrino waves at a distance $x$
is:
\beq{WW3a}
\phi_m(x)= (p_1 - p_2)\cdot x =
{{(p_1^2 - p_2^2)}\over{(p_1 + p_2)}}\cdot x  =
{{\Delta m^2}\over{2p}}\cdot x \approx
- \left({{\partial p}\over{\partial (m^2)}}\right)_{E}
\Delta m^2\cdot x
\eeq
where $\Delta m^2 \equiv m_2^2-m_1^2$, we have assumed the free space
relation between the masses, $m_i $ energy $E $ and momenta:
$~ ~ p_i^2 = E^2 - m_i^2$, noted that
$|m_2 - m_1| \ll p \equiv (1/2)(p_1 + p_2)$ and
kept terms only of first order in $m_2 - m_1$.
This result is seen to agree with eq. (\ref{YZY52}) obtained by the use of
handwaving A.
 
Thus we have a complete
solution to the oscillation problem and can give the neutrino flavor as a
function of the distance to the detector by examining the behavior of a single
energy eigenstate. Flavor-energy factorization enables the result to be
obtained without considering interference effects between different energy
eigenstates. All such interference is time dependent
and required to vanish at the source, where the flavor is time independent.
This time independence also holds at the detector as long as
there is significant overlap between the wave packets for different mass states.
The only information needed
to predict the neutrino oscillation wave length is the behavior
of a linear combination of the three mass eigenstates having
the same energy and different momenta.
Same energy and different momenta are relevant rather than vice versa
because the measurement is in space, not time, and flavor-time factorization
holds in a definite region in space.
 
We now note that this solution (\ref{WW3a}) enables a simple rigorous
justification of handwaving A to first order in the mass difference $m_2 - m_1$.
The standard relativistic energy-momentum relation gives the following relation
between the change in energy or momentum with mass when the other is fixed,
\beq{WW4b}
\left({{2E\partial E}\over{\partial (m^2)}}\right)_{p}
= - \left({{2p\partial p}\over{\partial (m^2)}}\right)_{E} = 1.
\eeq
Thus if
\beq{WW4g}
x ={{p}\over{E}} \cdot t
\eeq
\beq{WW4f}
\left({{\partial p}\over{\partial (m^2)}}\right)_{E}\cdot x =
\left({{E}\over{p}}\cdot{{\partial E}\over{\partial (m^2)}}\right)_{p}
\cdot x =
\left({{\partial E}\over{\partial (m^2)}}\right)_{p}
\cdot t
\eeq

\subsection {Generalization to cases with external fields}
 
The above treatment is now easily generalized to include cases where
flavor-independent external fields can modify the relation (\ref{WW1a}), but
where the mass eigenstates are not mixed by these fields, e.g. a gravitational
field. The relation between energy, momentum and mass is described by an
arbitrary dispersion relation
\beq{WW15}
f(E, p, m^2) = 0
\eeq
where the function $f$ can also be a slowly varying function of the distance
$x$. In that case, the momentum $p$ for fixed $E$ is also a slowly varying
function of $x$ and the x-dependence of the phase shift $\phi(x)$ is now
expressed by generalizing Eq. (\ref{WW3a})
to a differential equation
\beq{WW16a}
{{\partial^2  \phi(x)}\over{\partial x \partial (m^2)}}=
-\left({{\partial p}\over{\partial (m^2)}}\right)_{E}
= {{1}\over{v}}\cdot \left({{\partial E}\over{\partial (m^2)}}\right)_{p}
={{1}\over{v}}\cdot {{\partial^2  \phi(t)}\over{\partial t \partial (m^2)}}
\;, \qquad
v \equiv \left({{\partial E}\over{\partial p}}\right)_{(m^2)}
\eeq
where we note that the result can also be expressed in terms of the change
in energy with $m^2$ for constant momentum,
$\left({{\partial E}\over{\partial (m^2)}}\right)_{p}$,
instead of vice versa and the group velocity $v$, and can also be
expressed in terms of
the time-dependence of the phase shift measured at constant position.
We thus have generalized the justification (\ref{WW4f}-\ref{WW4g}) of
handwaving A to the case of a nontrivial dispersion relation by using the
group velocity of the wave.
 
Considerable confusion has arisen in the description of
flavor-oscillation experiments in quantum mechanics \cite{Kayser,NeutHJL},
with questions arising about time dependence and production
reactions \cite{GoldS}, defining precisely what exactly is
observed in an experiment \cite{Pnonexp}, and relations beween gedanken and
real experiments \cite{MMNIETO}.
Despite all these difficulties the expression (\ref{WW16a}) is seen to
provide an unambiguous value for the oscillation wave length in space and
also a rigorous recipe justifying Handwaving A for obtaining this oscillation
wave length from the period of oscillation calculated for a ``gedanken"
experiment which measures a gedanken oscillation in time. Note that the group
velocity and not the phase velocity enters into this relation.
 
The extension to propagation in a medium which mixes mass eigenstates e.g. by
the MSW effect is straightforward in principle, but more complicated in
practice and not considered here. The dispersion relation (\ref{WW15}) must
be generalized to be a nontrivial flavor-dependent $3 \times 3$ matrix
whose matrix elements depend upon $x$.
 
The exact form of the energy wave packet described by the function $g(E)$ is
irrelevant here. The components with different energies may be coherent
or incoherent, and they may be ``entangled" with other degrees of freedom of
the system. For example, for the case where a neutrino is produced together with
an electron in a weak decay the function $g(E)$ can also be a function
$g(\vec p_e,E)$ of the electron momentum as well as the neutrino energy.
The neutrino degrees of freedom observed at the detector will then be described
by a density matrix after the electron degrees of freedom have been properly
integrated out, taking into account any measurements on the electron. However,
none of these considerations can introduce a neutrino of the wrong flavor at the
position of the source.
 
Since the momenta $p_i$ are
energy-dependent the factorization does not hold at finite distance.
At very large values of $x$ the wave packet must separate into individual
wave packets with different masses traveling with different velocities
\cite{Kayser,Nus}.
However, for the conditions of a realistic oscillation experiment this
separation has barely begun and the overlap of the wave packets with different
masses is essentially 100\%. Under these conditions the flavor--energy
factorization introduced at the source is still an excellent approximation at
the detector. A detailed analysis of the separation process is given below.
 
The $\nu_e - \nu_\mu$ states with the same energy and different momenta are
relevant rather than vice versa
because the measurement is in space, not time, and flavor--time factorization
holds in a definite region in space.
 
In a realistic oscillation experiment the phase is important when the
oscillation length is of the same order as the distance between the
source and the detector.
In that case this
flavor-energy factorization holds over the entire distance between the source
and detector. The boundary condition then determines
the relative phase of components in the wave function with
different mass having the same energy and different momenta. Thus
any flavor oscillations observed as a function of the distance between the
source and the detector are
described by considering only the interference between
a given set of states having the same energy. All questions of coherence,
relative phases of components in the wave function with different energies and
possible entanglements with other degrees of freedom are thus avoided.
 
\section{Detailed analysis of A Pion Decay Experiment
$\pi \rightarrow \mu \nu$ }
 
We now consider an example of neutrino oscillations where the neutrinos are
produced by a $\pi \rightarrow \mu \nu $ decay from a pion brought to rest
in a beam dump and we consider the pion and muon wave functions in detail.
 
We first note that the pion is not free and is not at rest. It is still
interacting with the charged particles in the beam dump which have brought it
almost to rest. In the approximation where it is moving in the mean field
of the other charges, its wave function can be the ground state of motion in
this effective potential. In this case its energy $E_\pi$ is discrete and
uniquely defined, while its momentum will be just the zero-point or fermi
momentum described by a wave packet in momentum space,
\beq{WW51a}
\ket {\pi} = \int g(\vec p_{\pi}) d \vec p_{\pi} \ket {\pi(\vec p_{\pi}) }
\eeq
 
The decay is described by a weak interaction which commutes with the total
momentum of the system. Thus we can consider the decay of each individual
momentum component of eq.(\ref{WW51a}) separately. We assume that the width of
the wave packet in momentum space is sufficiently small so that we can neglect
the relativistic variation of the pion lifetime over the wave packet.
 
The energy, momentum and mass of the muon, denoted by $(E_\mu, p_\mu, m_\mu)$
and of the three mass eigenstates of the neutrino, denoted by $(E_i, p_i, m_i)$
where $i = 1,2,3$ are related by energy and momentum conservation:
\beq{WW51c}
E_i = E_\pi - E_\mu; ~ ~ ~ ~ ~ ~ \vec p_i = \vec p_\pi - \vec p_\mu
\eeq
\beq{WW51d}
E_i^2 = p_i^2 + m_i^2; ~ ~ ~ ~ ~ ~ E_\mu^2 = p_\mu^2 + m_\mu^2
\eeq
These relations differ from the corresponding relations for
the decay of a free pion because $E_\pi$ is a constant, independent of $p_\pi$.
It is determined by the binding potential and the energy change in the beam 
dump resulting from the removal of the pion. Since the final state of the beam 
dump is not measured, the results of the incoherent averaging over all final 
states is included by using the average energy change in the beam dump in 
$E_\pi$ in eq. (\ref{WW51c})
 
The final neutrino-muon wave function thus has the form:
\begin{eqnarray}
 \ket {(\mu,\nu)_f} & = & e^{-iE_{\pi}t}\cdot \int g(\vec p_{\pi}) d \vec p_{\pi}
\int d \vec p_{\mu}
\sum_{i=1}^3 \int d \vec p_i c_i e^{i\vec p_i \cdot \vec x_\nu}\cdot
\delta(E_\pi - E_\mu - E_i)
\cdot \nonumber \\
& &
\delta(\vec p_\pi - \vec p_\mu - \vec p_i)\cdot
\ket {\mu(\vec p_\mu),\nu_i(\vec p_i)}
\end{eqnarray}
where we have expressed the spatial dependence of the neutrino wave function
explicitly but left the spatial dependence of the muon wave function in the
wave function $\ket {\mu(\vec p_\mu),\nu_i(\vec p_i)}$,
$\ket {\nu_i}$ denote the three neutrino mass eigenstates and the
coefficients $c_i$ are left free and determined by the condition that the
neutrino must be a pure $\nu_{\mu}$ at the point $x_\nu = 0$ where the pion
decays.
 
The result of any experiment is obtained by taking the expectation value of
an operator $O_{exp}$ describing the measurement with the above wave function.
Since the muon and neutrino have separated by the time a
measurement is made, we assume that the operator factorizes into a product of
two operators $O_{\mu}$ and $ O_{\nu}$ acting on the muon and neutrino
respectively,
\beq{WW52a}
O_{exp} = O_{\mu}\cdot O_{\nu}
\eeq
We now assume that the muon operator $O_{\mu}$ commutes with the muon
momentum.
\beq{WW52b}
[O_{\mu},\vec p_\mu] = 0.
\eeq
This expression thus holds for any measurement in which the muon is not detected
as well as those where it is detected by an operator which commutes with its
momentum.
The experimental result is therefore given by the expression
\begin{eqnarray}
R_{exp} & = & \bra {\psi(\mu,\nu)} O_{exp} \ket{\psi(\mu,\nu)} =
\sum_{i=1}^3
\sum_{j=1}^3 \int \int \int \int \int d \vec p_{\mu} d \vec p_{\pi}
d \vec p'_{\pi} d \vec p_i d \vec p'_j g^*(\vec p_{\pi})
g(\vec p'_{\pi})
\cdot \nonumber \\
& &
 c^*_i \cdot c_j e^{i(\vec p_j - \vec p_i) \cdot \vec x_\nu}\cdot
\delta(E_\pi - E_\mu - E_i)\delta(E_i - E'_j)
\delta(\vec p_\pi - \vec p_\mu - \vec p_i)
\delta(\vec p'_\pi - \vec p_\mu - \vec p'_j) \nonumber \\
& &
\bra {\mu(\vec p_\mu),\nu_i(\vec p_\pi - \vec p_\mu)}O_{\mu}\cdot O_{\nu}
\ket {\mu(\vec p_\mu),\nu_j(\vec p'_\pi - \vec p'_\mu)}
\end{eqnarray}
 
We thus again obtain the result that the only interference terms that need
be considered are those between neutrino states having the same energy. The
crucial ingredient here is the unexpected relation between energy and momentum
of the stopped pion, which is not free. This is closely analogous to the
physics of the M\"ossbauer effect, where the relation between energy and
momentum for a nucleus bound in a lattice is crucially different from that for
a free nucleus. This resemblance between the treatment of recoil momentum
transfer in flavor oscillation phenomena and in the M\"ossbauer effect has been
pointed out\cite{LipQM} in the example of experiments measuring the $K_L - K_S$
mass difference by observing the regeneration of a $K_L$ beam as a function of
the distance between two regenerators. The coherence required depends upon the
impossibility of detecting the individual recoils of the two regenerators
resulting from the momentum transfer due to the mass difference.
 
\section{A Simple Pedagogical Neutrino Oscillation Puzzle}
 
\subsection{Statement of the Puzzle}
A pion at rest decays into a muon and neutrino. The neutrino oscillates between
electron neutrino and muon neutrino. We know everything and can calculate the
result of any neutrino oscillation experiment when the source is a pion at rest.
All factors of two are understood and the results agree with experiment.
 
How do we apply these results to a pion moving with relativistic velocity? A
naive picture of the conventional time dilatations and Lorentz contractions
occurring in moving systems suggests that the
oscillation period goes up, because of time dilatation, but the oscillation wave
length goes down because of the Lorentz contraction. Which wins? Is the
oscillation in time slowed down by the time dilatation? Is the oscillation in
space speeded up by the Loretnz contraction? What hapens in a real experiment
with Fermilab neutrinos? In a long baseline experiment?
 
Of course the real result is given above in eq. (\ref{WW3a}) and there is no
ambiguity. But what is wrong with the naive picture of time dilatations and
Lorentz contractions?
Note that this statement of the problem separates relativity from quantum
mechanics by assuming that the quantum mechanics is already solved in the
pion rest frame, and that only a Lorentz transformation to a moving frame is
needed.
 
\subsection{Pedestrian Solution to Puzzle}
 
Consider a $45^o$ mixing angle with a pion at rest and a detector at just the
right
distance so that it detects only electron neutrinos and no muon neutrinos. For
a qualitative picture of the physics, consider the Lorentz transformation to a
frame moving with velocity $v$, and assume that the pion decay and the neutrino
detection occur at the points $(x,t) = (0,0)$ and $(X,T)$, where we can as a
first approximation let $X=T$, with $c=1$ and assume that the velocity $v$ of
the frame is not too large. For a one-dimensional case we immediately obtain
 
\beq{WW55a}
(X,T)  \rightarrow (X',T') = {{X - vT, T - vX}\over{\sqrt{1-v^2}}} =
(X,T)\cdot \sqrt{\left[{{1-v}\over{1+v}}\right] }
\eeq
 
We now note that the neutrino momentum and energy (p.E) undergo the
transformation in the same approximation
\beq{WW55b}
(p,E)  \rightarrow (p',E') = {{p - vE, E - vp}\over{\sqrt{1-v^2}}} =
(p,E)\cdot \sqrt{\left[{{1-v}\over{1+v}}\right] }
\eeq
Thus
\beq{WW55c}
{{X'}\over{p'}} = {{X}\over{p}}
\eeq
 
So the observed oscillation wave length and period both decrease if the neutrino
is emitted backward and increase if the neutrino is emitted forward. The
backward emission is not relevant to realistic experiments. The naive pictures
are not relevant because the Lorentz contraction
always refers to two events occurring AT THE SAME TIME in each frame, and not
to the distance between THE SAME TWO EVENTS observed in different frames.
 
That both the wave length and period must vary in the same fashion is very
clear in this approximation where the motion is on the light cone which gives
X=T in all frames.
 
Thus the ratio $X/p$ is invariant and the
expression (\ref{WW3a}) for the relative phase of the two neutrino waves
holds also in a moving frame. Thus the result of the standard treatment is seen
to hold also for neutrinos emitted in the decay of a moving pion.
 
We now correct for the deviation of the velocity of the neutrinos from c
by writing
\beq{WW56a}
X = (p/E)T
\eeq
Thus
\beq{WW56b}
X  \rightarrow X' = {{X - vT}\over{\sqrt{1-v^2}}} =
 {{X[1 - v(E/p]}\over{\sqrt{1-v^2}}}
\eeq
and
\beq{WW56c}
p\rightarrow p= {{p - vE}\over{\sqrt{1-v^2}}} =
 {{p[1 - v(E/p]}\over{\sqrt{1-v^2}}}
\eeq
Thus the expression (\ref{WW55c}) holds for the general case and the
result of the standard treatment remains also when corrections for the
deviations of the neutrino velocity from c are taken into account.
 
\section {\bf Space and Time in Flavor Oscillations }
 
\subsection {Description in terms of time behavior}
 
\subsubsection{Fuzziness in Time}
 
In a neutrino oscillation experiment there must be uncertainties in order to
have coherence and oscillations. If we know that a neutrino has left a source
at time $t(s)$ and has arrived at the detector at a time $t(d)$, then we know
that its velocity is
\beq{YY1}
v = {{x}\over{t(d) - t(s)}}                         \eeq
where $x$ is the distance between source and detector.
We therefore know its mass and there are no oscillations.
 
In order to observe oscillations we cannot know exactly all the variables
appearing in eq. (\ref{YY1}). If oscillations are observed, there must be uncertainty
somewhere. It is easy to show that the major uncertainty must be in the time
$t(s)$ in which the neutrino is emitted from the source.
 
A detailed description of the time behavior and the need for fuzziness
in time is given in ref.\cite{GrossLip}. We summarize here the result
showing quantitatively the analog with the optical case.
 
If the mass eigenstate wave packets leave the source with their centers
together at $x=0$ the displacement between their centers at the point $x_d$ of
the detector is
\beq{QQ801}
\delta x_c = {{\delta v}\over{v}}\cdot x_d  \approx
 {{\delta p}\over{p}}\cdot x_d  =
{{\Delta m^2 }\over{2p^2}}\cdot x_d \,,\qquad
\delta v \equiv v_1 - v_2 \,,\qquad \delta p \equiv p_1 - p_2\,,
\eeq
where $\delta v$ and $\delta p$ denote the velocity and momentum differences
between the two mass eigenstates.
The neutrino masses are much smaller than their energies,
\beq{QQ802}
m_i^2 = E_i^2 - p_i^2 \ll p_i^2
\eeq
The neutrino can be detected at the detector when any point in the wave
packet passes $x_d$.
 
\subsubsection{Detailed description of time behavior and time overlaps}
 
Let $\ket{m_1}$ and $\ket{m_2}$ denote the two mass eigenstates and
$\theta$ denote the mixing angle defining the flavor eigenstates denoted by
$\ket{f_1}$ and $\ket{f_2}$ in terms of the mass eigenstates,
\beq{QQ807}
\ket{f_1} = \cos \theta \ket{m_1} + \sin \theta \ket{m_2} \,,\qquad
\ket{f_2} = \sin \theta \ket{m_1} - \cos  \theta \ket{m_2}\,,
\eeq
The wave function at the position of the detector at a time $t$ can be
written as a linear combination of the two mass eigenstates. We assume that
the the amplitudes denoted by $A(t)$ of the two wave packets are the same, but
that they are separated in time at the detector by the time interval
\beq{QQ808}
\tau_d = {{x_d}\over{v_2}} - {{x_d}\over{v_1}} \approx
{\delta v \over v^2} \cdot x_d
\approx {{\Delta m^2 }\over{2p^2 v}} \cdot x_d\,,
\eeq
The wave function at the detector can therefore be written
\beq{QQ806}
\ket{\Psi_d(t)} = e^{i \phi_o(t)}
\left[ \cos \theta A(t)  \ket{m_1} +
\sin \theta A(t + \tau_d) e^{i  \phi(\tau) }\ket{m_2}\right]\,,
\eeq
where $\phi_o(t)$ is an overall phase factor and
\beq{QQ814}
\phi(\tau)=p \delta x_c = p\,v\, \tau_d \approx {{\Delta m^2 }\over{2p}}x_d
\eeq
is the relative phase between the two mass eigenstates at the detector
The probability amplitudes and the relative probabilities that flavors
$f_1$ and $f_2$ are observed at the detector are then
\beq{QQ809}
\langle f_1 \ket{\Psi(t)} = e^{i \phi_o(t)}
\left[\cos^2 \theta A(t) e^{i  \phi(\tau) }+
\sin^2 \theta A(t+ \tau_d)  \right]\,,
\eeq
\beq{QQ810}
\langle f_2 \ket{\Psi(t)} = e^{i \phi_o(t)} \sin \theta \cos \theta
\left[ A(t) e^{i  \phi(\tau) } -
A(t+ \tau_d)  \right]\,.
\eeq
\beq{QQ811}
P(f_1,\tau_d) = \int dt |\langle f_1 \ket{\Psi(t)}|^2 =
 1 - {\sin^2 (2\theta) \over 2} \Big[1 - O(\tau_d) \cos \phi(\tau)\Big]\,,
\eeq
\beq{QQ812}
P(f_2,\tau_d) = \int dt |\langle f_2 \ket{\Psi(t)}|^2 =
 {\sin^2 (2\theta) \over 2} \Big[1 - O(\tau_d) \cos \phi(\tau) \Big]\,,
\eeq
where we have normalized the amplitudes and $O(\tau_d)$ is the time overlap
between the mass eigenstates,
\beq{QQ813}
\int dt |A(t)|^2 = 1 \,,\qquad
O(\tau_d) \equiv \int dt A(t + \tau_d) A(t)\,.
\eeq
We thus see how the standard result for neutrino oscillations arises for
the case where the overlap integral $O(\tau_d) \approx 1$ and how the
incoherent mixture of the two mass eigenstates is approached as
$O(\tau_d) \Rightarrow 0 $.
 
\subsection {When do Mass Eigenstate Wave Packets Separate? }
 
Suppose a wave packet is created which is a coherent linear combination of
two mass eigenstates, and the overlap of the two mass components is nearly
100\%. In time both wave packets will spread, and the centers will separate.
Will the separation between the centers of the packets be greater than the
spreading? Will there be an eventual spatial separation between the two
mass eigenstates? It is easy to see that in the extreme relativistic limit
the wave packets will separate; in the nonrelativistic limit they will not.
We simply need to calculate the velocities of the different components of
the packet.
 
Let $(\delta p)_W$ denote the momentum spread within each wave packet and
$(\delta p)_m$ denote the momentum difference between the components of the two
mass-eigenstate wave packets with the same energy.
 
The spread in velocity within a wave packet $(\delta v)_W$ is just the
difference in velocities $v = p/E$ for states with different momenta and the
same mass,
  \beq{ZZ2a}
 (\delta v)_W = {{\partial }\over{\partial p}}\cdot\left(
{{p}\over{E}}\right)_m \cdot (\delta p)_W =
{{(\delta p)_W }\over{E}} \cdot  {{m^2}\over{E^2}}
     \eeq     
The difference in velocity between components in two wave packets
 $(\delta w)_m$ with the same energy and different mass is just the
difference in velocities $v = p/E$ for states with different momenta and the
same energy,
  \beq{ZZ2b}
(\delta v)_m = {{\partial }\over{\partial p}}\cdot \left(
 {{p}\over{E}}\right)_E \cdot (\delta p)_m =
{{(\delta p)_m }\over{E}}      
     \eeq
The ratio of the spreading velocity to the separation velocity is then given by
  \beq{ZZ3}
{{(\delta v)_W}\over{(\delta v)_m}} = {{(\delta p)_W }\over
{(\delta p)_m }} \cdot  {{m^2}\over{E^2}}
         \eeq  
 
In the nonrelativistic limit where $E \approx m$ the ratio of the spreading
velocity to the separation velocity is just equal to the ratio of the momentum
spread in the wave function $(\delta p)_W$ to the momentum difference between
the two mass eigenstate wave packets. This will be much greater than unity if
there is to be appreciable overlap between the two wave packets in momentum
space.
 
  \beq{ZZ4}
     {{(\delta p)_W }\over {(\delta p)_m }} \gg 1
\eeq
Otherwise there will be no coherence and no spatial oscillations
observed. Thus in the nonrelativistic limit two wave packets which have an
appreciable overlap in momentum space will never separate.
 
In the relativistic case, the ratio of the spreading velocity to the
separation velocity is reduced by the factor $ {{m^2}\over{E^2}}$. This is
effectively zero in the extreme relativistic limit $ E \gg m $ relevant for
neutrino oscillations. Here the spreading velocity of the
wave packet is negligible and the wave packets will eventually separate.
     \beq{YY399}
{{m^2}\over{E^2}} \approx 0; ~ ~ ~
{{(\delta v)_W}\over{(\delta v)_m}} \ll 1
     \eeq
 
\subsection {At what distance is coherence lost?}
 
\subsubsection{The condition on the momentum spread in the wave packet}
 
Neutrino oscillations are always described in the relativistic limit and the
wave packets corresponding to different mass eigenstates will eventually
separate. Once they have separated they will arrive at a detector at different
separated time intervals. The detector will see two separated probability
amplitudes, each giving the probability that the detector will observe a given
mass eigenstate and all coherence between the different mass eigenstates will be
lost. The question then arises when and where this occurs; i.e. at what
distance from the source the coherence begin to be lost. We now examine two
different approaches to this problem and find that they give the same answer.
 
1. The centers of the wave packets move apart with the relative velocity
$(\delta v)_m$ given by eq. (\ref{ZZ2b}). Thus the separation $(\delta x)_m$ between
the wave packet centers after a time $t$ when the centers are at a mean distance
$x$ from the source is
\beq{WW5a}
(\delta x)_m=
(\delta v)_m \cdot t= (\delta v)_m \cdot {{x}\over{v}}=
- {{\Delta m^2}\over{2pE}} \cdot {{xE}\over{p}}=
- {{\Delta m^2}\over{2p^2}}\cdot x
\eeq
 
The wave packets will separate when this separation distance is comparable to
the length in space of the wave packet. The uncertainty principle suggests that
the length of the wave packet $(\delta x)_W $  satisfies the relation
\beq{WW6a}
(\delta x)_W \cdot (\delta p)_W \approx 1/2
\eeq
The ratio of the separation over the length is of order unity when
\beq{WW7a}
\left|{{(\delta x)_m}\over{(\delta x)_W}}\right|  \approx
\left| {{\Delta m^2}\over{p^2}}\right| \cdot (\delta p)_W
 \cdot x \approx 1 \eeq
 
2. Stodolsky\cite{Leo} has suggested that one need not refer to the time
development of the
wave packet, but only to the neutrino energy spectrum. With this approach we
note that the relative phase $\phi_m(x)$ between the two mass eigenstate waves
at a distance $x$ from the source depends upon the neutrino momentum $p_\nu$
as defined by the relation (\ref{WW3a}).
 
Coherence will be lost in the neighborhood of the distance $x$ where the
variation of the phase over the momentum range $(\delta p)_W$ within the wave
packet is of order unity.
For the case of two neutrinos with energy $E$ and mass eigenstates
$m_1$ and $m_2$ the condition that the relative phase variation
$|\delta \phi_m(x)|$ between the two neutrino waves is of order unity
\beq{WW4a}
|\delta \phi_m(x)| = \left|{{\partial \phi_m(x)}\over{\partial p_\nu}}\right|
\delta p_\nu \cdot x = \left| {{\Delta m^2}\over{2p_\nu^2}}\right|
(\delta p)_W \cdot x
\approx 1 \eeq
 
We find that the two approaches give the same condition for loss of coherence.
 
\subsubsection{Evaluation of the momentum spread in the wave packet}
 
The value of the momentum spread $(\delta p)_W$ in the wave packet depends upon
the production mechanism. However, we can immediately see that this can be
simply estimated for all experiments in which the initial state is either a
beam impinging on a solid target or a radioactive decay of a source in a solid.
The momentum of the initial target or radioactive nucleus has momentum
fluctuations resulting from its confinement in a lattice with a spacing of the
order of angstroms. These momentum fluctuations then appear in the neutrino
momentum spectrum as a result of conservation of four-momentum in the neutrino
production process. One immediately sees that the momentum fluctuations are
much larger than the momentum difference between the different mass eigenstates
having the same energy, and that therefore the neutrino state produced at the
source has full coherence between the different mass eigenstates.
 
The momentum spread $(\delta p)_W$ is easily calculated in any experiment where
the spread is a result of the momentum spread $\delta p_{nuc}$ of a nucleus in
the initial state. This is just the neutrino energy change produced
by the Lorentz transformation which changes the momentum of the active nucleus
from zero to the finite value $\delta p_{nuc}$. The four-momentum $(p,E)$ of the
nucleus is changed by this transformation from $(0,M_{nuc})$  to
$(\delta p_{nuc},E_{nuc})$, where where $M_{nuc}$ and $E_{nuc}$ denote the
mass of the nucleus and the energy of the nucleus with momentum
$\delta p_{nuc}$. The small velocity $v$ of this Lorentz transformation is
given to first order in $v$ by
\beq{QQ1}
v \approx  {{\delta p_{nuc}}\over{M_{nuc}}}
\eeq
The neutrino four-momentum is changed from $(p_\nu,p_\nu)$ to
$[p_\nu + (\delta p)_W,p_\nu + (\delta p_W)]$. Thus
\beq{QQ2}
(\delta p)_W = {{(1+v)}\over{\sqrt {1-v^2}}}\cdot p_\nu - p_\nu \approx
{{\delta p_{nuc}}\over{M_{nuc}}}  \cdot p_\nu
\eeq
to first order in $v$.
 Substituting this result into the coherence condition (\ref{WW4a}) gives
\beq{QQ3}
|\delta \phi_m(x)| = \left|{{\Delta m^2}\over{2p_\nu}}\right| \cdot
{{\delta p_{nuc}}\over{M_{nuc}}}  \cdot x
\approx 1 \eeq
This can be rewritten
\beq{QQ4}
x \approx
\left|{{4p_\nu\cdot M_{nuc}}\over{\Delta m^2}}\right| \cdot
\delta x_{nuc}
 \eeq
where $\delta x_{nuc} \approx 1/(2 \delta p_{nuc})$ denotes the quantum
fluctuations of the position of the nucleus.  This uncertainty principle
relation is an exact equality for the harmonic potential generally used
to describe binding in crystal lattices.
Because of the very different scales of the variables appearing in eq.
(\ref {QQ4}) we rewrite this relation expressing $x$ in kilometers,
$\delta x_{nuc}$ in Angstroms, $M_{nuc}$ in GeV, $p_\nu$ in MeV and $m$ in
electron volts.  In these units eq. (\ref {QQ4}) becomes
\beq{QQ5}
x(km) \approx 400 \cdot
\left|{{p_\nu(MeV)\cdot M_{nuc}(GeV)}\over{\Delta m(ev)^2}}\right|\cdot
\delta x_{nuc}(Angstroms)
 \eeq
This is seen to be a very large distance even for the case where the neutrino
originates from a solid where nuclei are confined to distances of the order of
Angstroms. For atmospheric and solar neutrinos, where the source is free to
move in distances many orders of magnitudes larger, the decoherence distance
will be even larger.
This calculation confirms the
result quoted in Kim and Pevner's book, chapter 9, that the coherence is lost
only at astronomical distances much larger than the size of the solar system
and that this coherence loss is relevant only for supernova neutrinos.
Note that the present derivation avoids making assumptions like those used
by Kim and Pevsner in which the neutrino is produced at time t=0, and which can
be questioned as shown below because of the uncertainty necessary for
coherence.
 
\section{\bf Space, Time, Relativity and Quantum Mechanics}
 
We now present a simple picture to guide intuition through all the arguments
about relativity, proper time, and the equivalence of space and time. In all
experiments the neutrino leaves the source as a wave packet which has a
finite length in space and time. If a detector is set up to detect the neutrino
at
a given point in space, the wave packet passes the detector during a finite time
interval. The probability of observing the neutrino at this point in space will
therefore have a statistical distribution in time given by the square of the
amplitude of the wave packet.
 
In principle, it is possible to measure time, rather than distance. This can
give a photographic record of the square of the wave packet in space at a
given instant of time. In principle it is possible to measure both the position
in space and the exact time for each detected neutrino event. The results can
be
presented as a scatter plot with space position and time plotted for each event.
The events for a given space position will show a time distribution over a
finite interval. The events for a given time will show a space
distribution over a finite interval. There is complete symmetry between space
and time, and there is a statistical distribution also of proper times.
 
How does one get physics out of these distributions? In practice it is only the
space position of the detected event that is measured, and it is known that
the probability of finding a neutrino with the wrong flavor at the source
must vanish. This determines the relative phase of the neutrino eigenstates as
they propagate through space. This is all the information needed to
give a unique interpretation for the results of any experiment.
 
There have been some suggestions that radioactive sources with long lifetimes
can introduce additional effects due to the long lifetime. Such effects have
been known and observed experimentally in electromagnetic transitions. However
the neutrino is a fermion, not a boson, and its emission must be accompanied
by the emission or absorption of another fermion. This change in the
environment is observable and ``collapses the wave function". If we
are considering a long-lived beta decay of a nucleus bound in an atom, the
nuclear lifetime is irrelevant for neutrino coherence because the nucleus
is interacting with the atom, and the atom knows when the charge of the
nucleus has changed and an electron or positron has been emitted together
with the neutrino.
 
The point has been repeatedly made by Stodolsky\cite{Leo} that the proper
formalism to treat neutrino oscillations is the
density matrix, because only in this way the unavoidable interactions with
the environment can be taken into account. This paper also points out that the
length in time of the wave packet is irrelevant.
 
\section{conclusions}
 
Flavor oscillations have been shown to be simply described in a wave picture,
very analogous to optical polarization rotations. The flavor eigenstates are
analogous to spin polarization eigenstates, and the neutrino oscillations
are describable as rotations in some abstract flavor-spin space.
 
The simplest description begins with the detector, which is located at a
definite position in space and which responds in a well-defined manner to the
arrival of some mixture of neutrino mass eigenstate waves. These individual
waves have traveled with different velocities from the source to the detector,
but have been shown to separate very slowly under practical conditions. Thus
there is almost a complete overlap at the detector except for neutrinos
arriving from distances much larger than the distance between the earth and the
sun; e.g. for neutrinos arriving from supernova.
 
The crucial parameters which determine the response
of the detector are the relative phases of the mass eigenstate waves at the
detector. These are determined by the initial conditions at the source and
by the propagation between source and detector. The propagation is
straightforward for free space and is well-defined for passage through
known external fields or media with well-defined properties; e.g. MSW effects.
The initial conditions at the source may be more complicated, depending upon
the particular reactions in which neutrinos are produced.
 
The fact
that all experiments in which oscillations can be measured involve sources
which are very small in comparison with the oscillation wave length
enables results to be easily obtained by using a universal boundary
condition: the probably of finding a particle with the wrong flavor at the
source must vanish. These results confirm the standard procedure of
calculating oscillations in time and converting a frequency in time to a
wave length in space by using the mean group velocity of the wave. That it
must be the group velocity has been shown rigorously for cases where the
neutrino is not free but may be subject to external fields like a
gravitational field.
 
The role of the quantum-mechanical uncertainty principle has been shown to
be crucial. Considerable care must be taken in using a particle picture with
well-defined times and momenta, rather than a wave picture with times and
momenta described by probability  amplitudes. Most published conclusions
regarding oscillations of recoil particles have been
shown\cite{goldman1,goldman2,okun1}  to be incorrect;
No such muon or $\Lambda$ oscillations should be observed.
 
\section{acknowledgments}
It is a pleasure to thank Leonid Burakovsky, Terry Goldman, Yuval Grossman,
Boris Kayser, Lev Okun and Leo Stodolsky for helpful discussions and
comments.

\end{document}